\documentstyle[11pt,aaspp4]{article}
\tighten

\input psfig.tex 

\def\check_mode#1{\ifmmode{#1}\else{$#1$}\fi}

\def\ddeg   {\check_mode{{\rlap.}^\circ}}

\begin{document}

\def\eg{{\it e.g. }}
\def\ie{{\it i.e. }}
\def\FIRAS{{FIRAS}}
\def\COBE{{\sl COBE\/}}
\def\specpaper{Mather \etal\ (1994)}
\def\calpaper{Fixsen \etal\ (1994b)}
\hfuzz=10pt \overfullrule=0pt

\def\blankline{\par\vskip \baselineskip}

\def\Received{\par\vskip 36 pt
        \centerline{\it Received  \vrule height -1 pt width 2.5 in
        depth 1.8 pt .}\par\blankline}

\def\refitem #1! #2! #3! #4;{\hang\noindent
    \hangindent 20pt\rm #1, \it #2, \bf #3, \rm #4.\par}
\def\bookref{\par\noindent\hangindent 20pt}
 
\raggedright

\def\wisk#1{\ifmmode{#1}\else{$#1$}\fi}
\def\lt     {\wisk{<}}
\def\gt     {\wisk{>}}
\def\um     {\wisk{{\rm \mu m\ }}}
\def\etal   {{\sl et~al.\ }}

\pretolerance=1000	

\title{The Spectrum of the CMB Anisotropy from the Combined
{\it COBE}\altaffilmark{1} FIRAS and DMR Observations}

\author{ D.J. Fixsen\altaffilmark{2,3},
         G. Hinshaw\altaffilmark{4},
	 C. L. Bennett\altaffilmark{4},
      \& J.C. Mather\altaffilmark{4} }

\altaffiltext{1}{The National Aeronautics and Space Administration/Goddard 
Space Flight Center (NASA/GSFC) is responsible for the design, development, 
and operation of the Cosmic Background Explorer ({\it COBE}).
Scientific guidance is provided by the {\it COBE} Science Working Group.
GSFC is also responsible for the development of the analysis software and
for the production of the mission data sets.}
\altaffiltext{2}{Hughes STX Corporation, 
                 Code 685, NASA/GSFC, 
                 Greenbelt MD 20771.}
\altaffiltext{3}{e-mail: fixsen@stars.gsfc.nasa.gov}
\altaffiltext{4}{Laboratory for Astronomy and Solar Physics,
                 Code 685, NASA/GSFC, 
                 Greenbelt MD 20771.}

\begin{abstract}

We analyze the Cosmic Microwave Background (CMB) anisotropy data from the 
independent {\it COBE} FIRAS and DMR observations. We extract the frequency 
spectrum of the FIRAS signal that has the spatial distribution seen 
by DMR and show that it is consistent with CMB temperature fluctuations in the  
radiation well into the Wien region of the spectrum.  Conversely, we form a 
map of the Planckian component of the sky temperature from FIRAS and show 
that it correlates with the DMR anisotropy map. The $rms$ fluctuations at
angular scales of 7$^\circ$ are 48$\pm 14~~\mu$K for the FIRAS data $vs$
35$\pm 2~~\mu$K for the DMR data and 31$\pm 6~~\mu$K for the combination 
(1 $\sigma$ uncertainties). The consistency of these data, from very different 
instruments with very different observing strategies, provide compelling 
support for the interpretation that the signal seen by DMR is, in fact, 
temperature anisotropy of cosmological origin.  The data also limit $rms$ 
fluctuations in the Compton $y$ parameter, observable via the 
Sunyaev-Zel'dovich effect, to $\Delta y < 3 \times 10^{-6}$ (95\% CL) 
on $7^{\circ}$ angular scales.
\end{abstract}


\section{Introduction} 

The DMR (Differential Microwave Radiometers) instrument aboard the {\it COBE} 
satellite (Boggess \etal\ 1992 and references therein, Smoot \etal\ 1990) was 
designed to measure anisotropy in the CMB temperature on $7^{\circ}$ angular 
scales.  The detection of anisotropy (Smoot \etal\ 1992, Bennett \etal\ 1996 
and references therein) provides important information on conditions in the 
early universe.  It is widely interpreted that the signal seen by DMR is of 
cosmological, and possibly primordial, origin.  As such, it provides key 
information on gravitational potential fluctuations in the early universe 
that presumably served as the seeds for the formation of large scale 
structure.  The primary evidence that the anisotropy is cosmological arises 
from a spectral analysis of the 3 DMR frequency channels at 31.5, 53, and 90 
GHz.  Cosmological anisotropy is predicted to have a Planckian spectrum of the 
form $I(\nu) = B(\nu,T_0) + \Delta T \partial B(\nu,T)/\partial T$, where 
$T_0 = 2.728$ K is the mean temperature of the CMB (Fixsen \etal\ 1996), 
$B(\nu,T)$ is the Planck function at temperature $T$, and $\Delta T = T-T_0$ 
is the temperature anisotropy in a given direction in the sky.  The DMR 
anisotropy data are consistent with a Planck spectrum and are inconsistent 
with any known Galactic emission (Hinshaw \etal\ 1996, Banday \etal\ 1997).  
However, the frequency range covered by DMR is relatively limited, and 
observations of the anisotropy into the Wien region of the spectrum would 
greatly strengthen the view that the anisotropy is due to temperature 
fluctuations in the CMB.  Ganga \etal\ (1993) cross-correlated the FIRS 
balloon-based anisotropy map with the DMR data and found a significantly 
correlated signal at 170 GHz.  To date, this has been the only direct test of 
the DMR at frequencies approaching the Wien region of the spectrum.

The FIRAS (Far Infrared Absolute Spectrophotometer) instrument aboard the 
\COBE\ satellite (Mather \etal\ 1993) was designed to  measure the spectrum 
of the CMB. The FIRAS observations demonstrate that the spectrum of the 
uniform (monopole) component of the CMB is Planckian to 50 ppm of the peak 
radiation intensity (Mather \etal\ 1990, Mather \etal\ 1994, Fixsen \etal\ 
1996).  Moreover, the CMB dipole exhibits a Planck spectrum ($\partial B /
\partial T$) to high precision (Fixsen \etal\ 1994a, Fixsen \etal\ 1996).  
In this paper we assess the spectrum and spatial distribution of the 
higher-order anisotropy in the FIRAS data.

The measurement of the anisotropy spectrum is difficult because the 
intrinsic anisotropy is weak, roughly one part in $10^5$ of the intrinsic CMB 
brightness, and near the limits of the FIRAS sensitivity and stability, {\it 
ie.} it is difficult to make absolute measurements of the CMB that are 
stable to 0.001\% over the course of a year of observations.  In addition, as 
one approaches the Wien region of the spectrum the signal becomes dominated by 
emission from interstellar dust.  To overcome these difficulties our analysis 
employs data from all three COBE instruments: the FIRAS data provide the 
spectrum, the DMR data provide the spatial distribution of the anisotropy, in 
effect telling us how to average the FIRAS spectral data, while the longest 
wavelength channels from the Diffuse InfraRed Background Experiment (DIRBE) 
are used to characterize the dust that would otherwise confuse the 
measurement.

We have analyzed the FIRAS data in two complementary ways.  First, the DMR and 
DIRBE maps are used as spatial templates to which the FIRAS data are fit.  The 
result is a single mean spectrum per map of the emission traced by that map.
The mean spectrum corresponding to the DMR anisotropy may be compared to the 
Planck spectrum given above.  Second, the FIRAS data are fit to three spectral 
functions to obtain three corresponding sky maps.  One of the spectra is taken 
to be the Planck function with the pixel temperature as a free parameter.  The 
other two are designed to model the spectrum of the dust with intensity and 
temperature as the free parameters.  The map of the (Planckian) CMB 
temperature may then be compared to the DMR anisotropy map using a 
cross-correlation analysis.

\section{The Data}

The low frequency FIRAS data consist of spectra between 2 and 20 cm$^{-1}$ (0.5 cm to 500 
$\mu$m wavelength) in each of 6068 pixels on the sky (there are 6144 pixels in 
the full sky).  They were calibrated using the method described in \calpaper, 
with the improvements noted in Fixsen \etal\ 1996. A weighted average of
all of the low frequency FIRAS data was used.
The DMR anisotropy map we use is a modification of the ``Correlation" 
technique reduced-Galaxy map described in Kogut et al. (1996).  It is a 
weighted average of all six DMR channel maps after correcting each for 
Galactic emission as traced by the Haslam 408 MHz radio map, the DIRBE 140 
$\mu$m map, {\it and} the DIRBE 240 $\mu$m map.  Our modification entails 
fitting the two DIRBE templates simultaneously which parallels our treatment 
of Galactic emission in the FIRAS data, described below.  We have also 
subtracted a best-fit monopole and dipole from the map, for $|b| \gt 
20^\circ$, prior to our analysis.  The long wavelength DIRBE data was 
corrected for low-level emission from interplanetary (zodiacal) dust by 
interpolating to the response at a solar elongation of $90^\circ$ 
corresponding to the FIRAS pointing.  All of the data sets were pixelized in 
ecliptic coordinates.

Both the FIRAS and DMR data have $7^{\circ}$ angular resolution, as measured 
by their full width at half maximum (FWHM), but the detailed beam profiles are 
different.  The DMR beam is approximately Gaussian (Wright \etal\ 1994), while 
the FIRAS beam more closely resembles a top hat profile.
Prior to fitting the higher resolution DIRBE (and 
Haslam) data to the FIRAS and DMR data we have convolved each with the FIRAS 
and DMR beam profiles, respectively.  Additionally, prior to comparing the 
FIRAS and DMR anisotropy data to each other, we have convolved the FIRAS data 
with a $7^{\circ}$ FWHM Gaussian.  We estimate that the resulting beam window 
functions agree to within 5\% to multipole order $\ell = 20$.

\section{Spectral Analysis}

The FIRAS sky spectra are a function of frequency and position: 
$S(\ell,b;\nu)$ where $\ell$ and $b$ are Galactic coordinates and $\nu$ is 
frequency. Since the data are binned into discrete pixels, $p$, and 
frequencies, $\nu$, we can also write the FIRAS data as $S_{p\nu}$.  Prior to 
investigating CMB anisotropy, we model the dominant emission components in the 
data by assuming the frequency and spatial dependences are separable, as 
follows
\begin{equation}
M_{p\nu} = u_\nu + D_p d_\nu + G_p g_\nu + H_p h_\nu
\end{equation}
where $M$ denotes our model of the data $S$.  The CMB monopole is represented 
by the uniform spectrum $u_\nu$; the CMB dipole anisotropy is represented by 
the spatial distribution $D_p$ and the spectrum $d_\nu$. Galactic emission is 
modeled using two spatial templates, $G_p$ and $H_p$, and two spectra $g_\nu$ 
and $h_\nu$.  In this analysis, we fix the spatial templates above and perform 
a least-squares fit of the data to determine the spectra, $u$, $d$, $g$, and 
$h$ at each frequency.

The spatial templates $D_p$, $G_p$, and $H_p$ are specified as follows.  The 
dipole is given by $D_p = \cos \theta$, where $\theta$ is the angle between 
the direction of observation and the dipole direction, $(\ell,b)$ = 
$(264.26^\circ,+48.22^\circ)$ (Bennett \etal\ 1996).  For the two Galactic 
templates we use the 240 $\mu$m and 140 $\mu$m DIRBE maps convolved to the 
FIRAS beam profile.  The use of two templates simultaneously allows for modest 
spatial variations in the Galactic dust spectrum and for a possible far 
infrared background.  To give the derived spectra common units of MJy 
sr$^{-1}$, we normalize the spatial templates to be dimensionless. The dipole 
map is scaled to range from $-1$ to $+1$.  The Galaxy maps are scaled such that 
the average value over the polar caps ($|b| \gt 60^\circ$) is 1.07, 
corresponding to the mean of csc$|b|$ over the same region.  The resulting 
Galaxy spectrum roughly gives the intensity at the Galactic poles.

For the fit we define $\chi^2$ to be
\begin{equation}
\chi_{\nu}^2 = \sum_{pp'}[S_{p\nu}-M_{p\nu}(u,d,g,h)]~W_{pp'}~
[S_{p'\nu}-M_{p'\nu}(u,d,g,h)]/\sigma_\nu^2
\end{equation}
where $p,~p'$ are sky pixel indices, $W$ is the inverse of the pixel-pixel 
covariance of the FIRAS data, including correlations introduced by the 
calibration and ``de-striping" processes, and $\sigma_\nu$ is the factorable 
uncertainty of the spectrum at frequency $\nu$, including ``glitch" effects 
(see Fixsen \etal\ 1996).  This $\chi^2$ is minimized independently at each 
frequency to determine the mean spectra $u$, $d$, $g$ and $h$.  It is 
important to note that this fit makes no a priori assumptions about the forms 
of these spectra; only their spatial distributions are posited. Note also that 
$\sigma^2_{\nu}$ does not affect the best-fit intensity at each frequency, 
only our estimate of the uncertainty.  The FIRAS data have been processed in 
such a way as to maintain the separability of the positional and the spectral 
uncertainties. 

For the discussion below we have excluded data with $|b| \lt 20^\circ$ 
although other Galactic cuts were examined. The derived spectra are shown in 
Fig 1.  The monopole and dipole spectra have been discussed extensively in 
Fixsen \etal\ (1996); this work does not alter any of their conclusions.
The Galactic spectra, Fig. 1c, are the shape one would expect for dust at 
$\sim15$ K.  The negative spectrum for the 140 $\mu$m map is a natural 
consequence of a frequency extrapolation that does not account for a varying 
dust temperature.  Indeed, if the dust temperature were constant, the 
(normalized) 240 $\mu$m and 140 $\mu$m maps would be identical and the fit 
would be degenerate. To study the Galactic dust emission one should look at 
the higher frequencies and lower latitudes; here we are only interested in 
removing the dust emission.

The FIRAS residual spectra, $R_{p\nu}=S_{p\nu}-M_{p\nu}$, now contain only CMB 
anisotropy, noise, and unmodeled signals and/or systematic effects.  We 
convolve the residual spectra with a 7$^\circ$ FWHM Gaussian beam to 
approximate the beam response of the DMR map.  This yields smoothed spectra, 
$F_{p\nu}$, with a covariance matrix $E_{pp'}\sigma^2_{\nu}$.  There are 
4004 pixels ($\sim 2.6^{\circ}$ square) remaining in the data after applying 
the 20$^{\circ}$ Galaxy cut and excising the pixels not observed by FIRAS.  
We normalize the DMR map to roughly unit $rms$ by dividing it by 35 $\mu$K, 
the estimated $rms$ of the anisotropy (Banday \etal\ 1997).  This allows us 
to compare the derived anisotropy spectrum to the previously derived spectra 
using consistent units.  We denote the normalized map $A_p$.  In accounting 
for the noise in both the FIRAS and DMR data, the best-fit spectrum, $a_{\nu}$, 
is given by
\begin{equation}
a_\nu = \sum_{pp'}(A_p U_{pp'} F_{p'\nu})/
\sum_{pp'}(A_p U_{pp'} A_p - U_{pp'}C_{pp'}),
\end{equation}
where $U$ is any symmetric matrix (see below), and $C$ is the covariance 
matrix of the DMR map which we take to be diagonal, $C_{pp'} = \sigma^2_p 
\delta_{pp'}$ (here $\sigma_p$ is the noise per pixel in the DMR map).  The 
second term in the denominator is a bias correction that accounts for the fact 
that noise in the template $A_p$ biases the first term.  Any symmetric matrix 
$U$ will give an unbiased estimate of the anisotropy spectrum; we use the 
inverse of an effective combined diagonal FIRAS and DMR matrix.  
The full covariance matrix is given by 
$E_{pp'}\sigma^2_{\nu\nu'} + (\partial B_{\nu}/\partial T)^2 C_{pp'}$. 
We define $U$ to be $(diagonal(E) + \alpha C)^{-1}$ with 
$\alpha = \sum_\nu (\partial B_{\nu}/\partial T)^2/\sigma_\nu^2$ 
for all frequencies.  This keeps the calculation simple and is reasonable 
since $E$ dominates the covariance as the FIRAS uncertainties are larger than 
the DMR uncertainties.

The anisotropy spectrum is shown in Fig. 1d. There is a significant detection 
of signal correlated with the DMR anisotropy: the $\chi^2$ for zero signal is 
75 for 43 d.o.f., while a  $(\partial B_{\nu}/\partial T)$ spectrum normalized 
to the DMR results give a $\chi^2$ of 48.  {\it This strongly suggests that 
the anisotropy observed by DMR, and corroborated by FIRAS, is due to 
temperature variations in the CMB.}  

We have also simultaneously fit all 5 maps (1, $D_p$, $G_p$, $H_p$, $A_p$) and 
have applied Galactic cuts between $10^\circ$ and $40^\circ$.  None of these 
cases show significantly different results, although with a Galactic cut of 
only $5^\circ$ there is evidence for Galactic contamination in the anisotropy 
spectrum.  We emphasize that best-fit Galactic templates have been subtracted 
from both the DMR and FIRAS maps, and that the same two Galactic templates 
were used on each.  The effects of the FIRAS gain variations and other 
nonlinear effects are entirely negligible because of the small amplitude of 
the anisotropy.  Most instrumental systematic effects are suppressed in the 
analysis because of the differences in the DMR and FIRAS instruments and scan 
patterns.

Figure 2 shows the anisotropy spectrum along with the {\it predicted} Planck 
spectrum based on the DMR data alone.  The data in this figure were binned 
coarsely in frequency for display, but our analysis of the spectrum was 
performed on the original data.  The data at different frequencies are mildly 
($\sim 10\%$) correlated, which we account for in our spectral analysis.
If we {\it fit} the FIRAS spectrum to a Planck spectrum, 
$\alpha ~\partial B_\nu / \partial T$,  we find correlated $rms$ temperature 
fluctuations of $\alpha = 28 \pm 5~\mu$K compared to the DMR value of 
$35 \pm 2~\mu$K (Banday \etal\ 1997).  We can further analyze the spectrum 
$a_\nu$ to search for evidence of spectral distortions due to the 
Sunyaev-Zel'dovich (SZ) effect, or for signs of residual Galactic 
contamination.  We fit $a_\nu$ to a model of the form 
$\alpha ~\partial B_\nu / \partial T + \beta ~(-1/2T_0) \partial 
B_\nu / \partial y + \gamma ~\nu^{\alpha_D} B_\nu(T_D)$, where $\alpha_D$ and 
$T_D$ are the dust emissivity and temperature respectively.  Since the DMR map 
was normalized to have unit $rms$, $\alpha$ corresponds to the $rms$ level of 
Planckian temperature anisotropy, $\beta$ corresponds to the $rms$ level of 
Sunyaev-Zel'dovich temperature anisotropy in the Rayleigh-Jeans limit, and 
$\gamma$ corresponds to the $rms$ level of dust emission relative to the 
emission at the Galactic poles.  Fixing $\alpha_D = 2$ and $T_D = 13$K, we 
find ($\alpha,\beta,\gamma$) = ($31 \pm 6~\mu$K, $5 \pm 11~\mu$K, $-.036 
\pm .017$).  The value of $\alpha$ is consistent with the measured DMR $rms$ 
of $35 \pm 2$ $\mu$K.  There is no evidence for Sunyaev-Zel'dovich anisotropy, 
though the fit does indicate a small amount of Galactic leakage.

We can also combine the 3 DMR single-frequency $rms$ fluctuation data at 
31.5, 53 and 90 GHz with the spectrum $a_{\nu}$ to form a single spectrum.  
These data are fit for $\alpha$, $\beta$ and $\gamma$ giving ($\alpha,\beta,
\gamma$) = ($32 \pm 4~\mu$K, $-2 \pm 5~\mu$K, $-.031\pm .015$), with 
$\alpha_D = 2$ and $T_D = 13 $K.
Treating the SZ coefficient as an upper limit, and assuming the SZ signal is 
uncorrelated with the primary anisotropy, we conclude that not more than 
10/35, or 30\%, of the DMR variance could be due to distortions from the 
Sunyaev-Zel'dovich effect at 95\% CL.  This corresponds to an upper limit of 
$\Delta T = \sqrt{10/35} \cdot 35$ $\mu$K = 18 $\mu$K, or $\Delta y = 
3 \times 10^{-6}$ for $rms$ fluctuations on a 7$^{\circ}$ angular scale, 
consistent with the limit on a uniform SZ distortion of $|y|\lt 15 \times 
10^{-6}$ (Mather \etal\ 1994).  Banday \etal\ (1996) cross-correlated 
the 4-year DMR maps with a variety of extra-galactic sources including Abell 
clusters and HEAO X-Ray data and found no evidence for correlated emission.  
They placed a 95\% CL upper limit of $\Delta y < 1 \times 10^{-6}$ for hot gas 
correlated with those particular sources.  The limits obtained here are higher 
in amplitude, but make no assumption about the spatial distribution of the gas 
and are based on direct measurements of the spectrum in the Wien region where 
the SZ distortion is manifest.  The limit on the dust fluctuations that are 
traced by DMR but {\it not} by DIRBE depend on the spectral properties of the 
assumed dust: we have fit $\alpha$, $\beta$ and $\gamma$ for a variety of 
dust properties and have computed the corresponding upper limits on $rms$ 
fluctuations at 90 GHz.  With $\alpha_D = 1$ we find $\Delta T_{rms} <$ 2.0, 
0.4, 0.3 $\mu$K for $T_D$ = 5, 10, 15 K respectively, and with $\alpha_D = 2$
we find $\Delta T_{rms} <$ 0.4, 0.1, 0.05 $\mu$K for $T_D$ = 5, 10, 15 K 
respectively.

\section{Correlation Analysis}

To complement the above spectral analysis, we now assume the spectral form of 
the CMB anisotropy in the FIRAS data and derive a map of the CMB temperature.  
The resulting map is then compared to the DMR map using a cross correlation 
analysis.  We assess the statistical consistency of the two anisotropy maps 
with the aid of Monte Carlo simulations.

To obtain a map of the CMB temperature, the FIRAS sky spectra are 
simultaneously fit to three spectral forms to derive three corresponding maps.  
As before, we model the data as
\begin{equation}
M_{p\nu} = B_\nu(T_0) + A_p\,\partial B_\nu/dT|_{T_0} + G_p\,g_\nu + H_p\,h_\nu
\end{equation}
where $T_0 = 2.728$ K is the mean CMB temperature.  The first two terms 
represent a Planck CMB signal with temperature $T_0 + A_p$, and the 
last two terms represent Galactic emission with two independent dust spectra, 
$g$ and $h$. Here we take $g = \nu^2B_\nu(13K)$ and $h = \nu^2 \partial B_\nu 
/ \partial T|_{13K}$.  (There is little difference in the anisotropy map if 
we use a different dust spectrum, as long as it allows for temperature 
variation across the sky.)  The spectral data are fit to this model 
independently in each pixel resulting in maps $A$, $G$, and $H$.  The Galactic 
maps are of limited interest since the fits use only the low frequency FIRAS 
data.  The anisotropy map, $A_p$, which has units of thermodynamic 
temperature, is dominated by the dipole variation in the CMB.  The agreement 
between the FIRAS dipole and the DMR dipole has already been established 
(Fixsen \etal\ 1996; Bennett \etal\ 1996).  The fits here confirm that 
agreement and verify that the relative calibration of the DMR and FIRAS 
instruments agree to within 0.5\%, consistent with the individual calibration 
uncertainties in each instrument.  In our cross-correlation analysis we remove 
the dipole from the maps by fitting for a dipole component on the portion of 
the sky that is being analyzed.  Also, as in the spectral analysis, we smooth 
the FIRAS anisotropy map, $A_p$, with a 7$^{\circ}$ FWHM Gaussian to 
approximate the response of the DMR beam.  The 2-point cross-correlation 
function is the average product of all pixel temperatures with a fixed angular 
separation: $C(\alpha) = \sum_{i,j} w^F_i w^D_j \delta T^F_i \delta T^D_j / 
\sum_{i,j} w^F_i w^D_j$ where the sum is over all pixel pairs ($i,j$) 
separated by $\alpha$, $w^F_i$ and $w^D_j$ are pixel weights appropriate to 
the FIRAS and DMR maps (to minimize noise, we adopt inverse variance 
weighting), and $\delta T^F_i$ and $\delta T^D_j$ are the temperatures in 
pixels $i$ and $j$ of the FIRAS and DMR maps respectively.  We bin the 
correlation data into 71 angular separation bins of width $2\ddeg6$.  The data 
are analyzed with Galactic cuts from 10$^{\circ}$ to 40$^{\circ}$ to determine 
the extent to which Galactic emission is affecting the results.

The 2-point functions obtained from the maps are shown in Figure 3.  We plot 
the auto-correlation functions of the FIRAS and DMR maps, and the cross 
correlation between the two.  The error bar attached to each point represents 
the $rms$ due to instrument noise and to modeled systematic effects in the 
FIRAS anisotropy map.  The plot demonstrates that:  1) the cross correlation 
of the FIRAS and DMR maps is consistent with the DMR auto-correlation 
function, and 2) The FIRAS auto-correlation function exhibits structure that 
is not present in the DMR map. The latter point is not surprising: signals of 
$\sim 50~\mu$K were expected.  The ``de-striping" process in the FIRAS data 
can leave artifacts of this size since the overall mission temperature 
calibration drift is a factor of $\sim$100 larger than the artifacts seen here.
Fortunately any residual instrumental effects in the 
FIRAS anisotropy map are uncorrelated with the DMR and are less than the 
statistical noise per pixel, thus they should only contribute modest additional 
noise to the cross-correlation function.  The residual stripes in the map may 
be described by a pixel-pixel covariance matrix.  Our Monte Carlo based error 
analysis includes both instrument noise and residual stripes as errors in the 
FIRAS map.

We test the consistency of the cross-correlation against the DMR 
auto-correlation quantitatively as follows.  The statistic for the test is 
defined as $\chi^2 = {\bf\Delta C}^T \cdot {\bf M^{-1}} \cdot {\bf \Delta C}$ 
where ${\bf \Delta C}$ is the observed difference between a pair of 2-point 
functions, and ${\bf M} = \langle ({\bf \Delta C})({\bf \Delta C})^T \rangle$ 
is the covariance matrix computed from simulations.  Specifically ${\bf \Delta 
C}$ has elements $\Delta C_a = C^{(FD)}_a - C^{(DD)}_a$ where $a$ denotes an 
angular separation bin, and $C^{(FD)}$ and $C^{(DD)}$ denote the cross- and 
auto-correlation functions respectively.  We generate an ensemble of 2-point 
functions for computing the covariance as follows: for each realization we 
compute a scale-invariant power-law sky map with a quadrupole normalization of 
18 $\mu$K, and filter it through the window function appropriate to each 
instrument's beam profile.  We generate 2 noise maps, one per instrument, with 
appropriate noise level and coverage (Bennett et al. 1996), and, in the case 
of FIRAS, one additional stripe map.  Each realization of the FIRAS signal + 
noise + stripes is smoothed with the 7$^{\circ}$ FWHM Gaussian.  The result is 
two maps with a common CMB signal, but different instrumental properties.  We 
then compute the auto-correlation of each map, and the cross correlation of 
the pair to form an ensemble of simulated 2-point functions.  From this we 
compute an ensemble of differences, ${\bf \Delta C}$, and their covariance, 
{\bf M}.

The first column of Table 1 shows the value of $\chi^2$ we obtain for the 
difference between the cross-correlation and the DMR auto-correlation for two 
Galaxy cuts.  This column also shows the fraction of simulated 2-point 
functions that had a larger difference, as measured by this statistic.  
Clearly, by this measure, there is substantial agreement between the two maps.  
Another measure of consistency is the quadrupole normalization of a 
scale-invariant signal fit to the auto and cross-correlation functions, using 
the method described in Hinshaw et al. (1996).  These results are also given 
in Table 1.  The normalization derived from the DMR auto-correlation, denoted 
$Q^{(DD)}$, is consistent with the value given in Hinshaw et al. (1996).  The 
normalization derived from the FIRAS-DMR cross-correlation, $Q^{(FD)}$, is 
consistent with the DMR normalization while it is inconsistent with zero, 
which rejects the null hypothesis that the two maps are uncorrelated.  Since 
the error associated with each determination of $Q$ in this method includes a 
contribution from cosmic variance, we have also quoted results for the 
normalization difference, $\Delta Q \equiv Q^{(FD)} - Q^{(DD)}$, for which the 
error depends only on instrument noise and systematic effects.  The 
normalization inferred from the cross correlation function is lower than the 
DMR normalization by roughly 1.5 standard deviations, which is consistent 
with the results obtained from the spectral analysis.

\section{Conclusions}

The mean FIRAS spectrum of the anisotropy observed by DMR is consistent 
with the Planck form, $\partial B / \partial T$, expected of CMB anisotropy.  
Conversely, the spatial distribution of the FIRAS data fit to a Planck 
spectrum agrees well with the DMR anisotropy.  Taken together, these results 
provide direct observational support for the widely held view that the 
signal first detected by the DMR is, in fact, temperature anisotropy in the 
CMB.

Observations of the dipole anisotropy corroborate the relative calibration of 
the two experiments to within 0.5\%.  We have used the anisotropy spectrum to 
place limits on the fraction of the DMR signal that could be due to 
Comptonization by hot gas via the Sunyaev-Zel'dovich effect.  We find that no 
more than 18 $\mu$K of the signal seen by DMR could be due to this mechanism 
at 95\% CL.  This corresponds to an upper limit on $rms$ fluctuations in the 
Compton $y$ parameter of $\Delta y < 3 \times 10^{-6}$ on $7^{\circ}$ angular 
scales, and rules out models in which the anisotropy is primarily due to the 
SZ effect.  We also limit fluctuations in thermal dust emission that are 
traced by DMR but {\it not} by DIRBE to less than 2 $\mu$K at 90 GHz.

We thank K. Jensen and J. Gales for many programs we used to process and 
analyze the data, J. Weiland for help in wrestling with the DIRBE data, and 
A.J. Banday, K. G\'orski, A. Kogut, S.H. Moseley, D. Spergel, R. Weiss, 
D. Wilkinson and E. Wright for helpful discussions.  This work was supported 
by the Office of Space Sciences at NASA Headquarters.

\clearpage
\begin{deluxetable}{clllll}
\tablewidth{5.5in}
\tablecaption{FIRAS-DMR Cross-Correlation Statistics}
\tablehead{ \colhead{Galaxy cut} &
            \colhead{$\chi^2$}   &
            \colhead{$Q^{(DD)}$} &
            \colhead{$Q^{(FD)}$} &
            \colhead{$\Delta Q$\tablenotemark{a}} &
            \colhead{$Q^{(FF)}$} \nl
            \colhead{(degrees)}  &
            \colhead{}           &
            \colhead{($\mu$K)}   &
            \colhead{($\mu$K)}   &
            \colhead{($\mu$K)}   &
            \colhead{($\mu$K)}   }
\startdata
20 & 65 (54\%) & $17.9 \pm 1.6$ & $14.4 \pm 2.4$ & $-3.5 \pm 2.2$ 
               & $24.8 \pm 7.1$ \nl
30 & 66 (53\%) & $16.5 \pm 1.8$ & $15.1 \pm 2.8$ & $-1.4 \pm 2.4$ 
               & $24.8 \pm 7.8$ \nl
\enddata
\tablenotetext{a}{$\Delta Q \equiv Q^{(FD)} - Q^{(DD)}$.  The quoted 
uncertainty on $\Delta Q$ accounts for correlations between $Q^{(FD)}$ and 
$Q^{(DD)}$.}
\label{cc_stat_table}
\end{deluxetable}

\clearpage
\typeout{FIGURE CAPTIONS}

\begin{figure}[t]
\caption{Spectra resulting from the fit to fixed spatial templates with 
1 $\sigma$ errors.  a) The monopole spectrum with errors magnified by a 
factor of 200, the curve is a 2.728 K Planck spectrum. b) The dipole spectrum, 
the curve is a 3.36 mK differential Planck spectrum. c) The Galactic spectra, 
the dashed and dotted lines are the fits to the 240 and 140 $\mu$m maps 
respectively, and the points are the sum. d) The anisotropy spectrum, the 
curve is the 35 $\mu$K differential Planck spectrum predicted by DMR and the 
dashed line is 3.6\% of the Galactic spectrum in c. }
\label{fspectra}
\end{figure}

\begin{figure}
\caption{Mean anisotropy spectrum, as in Fig. 1d, with the points averaged in 
frequency, and with the fitted 3.6\% Galactic residual spectrum removed.  The 
uncertainties are 1 $\sigma$.  The solid line is the differential Planck 
spectrum predicted by DMR, and the dashed line with error band is the best-fit 
Planck spectrum.}
\label{fcuts}
\end{figure}

\begin{figure}
\caption{2-point correlation functions obtained from the FIRAS and DMR maps.
a) The FIRAS (squares) and DMR (diamonds) auto-correlation functions.  The 
uncertainties in the FIRAS data are dominated by a quadrupole and have been 
factored for display: the solid curve shows a 1 $\sigma$ fluctuation in the 
(mostly quadrupole) mode, the rest of the uncertainty is indicated by the 
partially correlated error bars. The data themselves have not had a quadrupole 
subtracted.  b) The FIRAS-DMR cross correlation function (squares) and the 
DMR auto-correlation function (diamonds).  The error bars represent instrument 
noise and, in the case of FIRAS, modeled systematic effects.  c) The 
difference between the FIRAS-DMR cross-correlation and the DMR 
auto-correlation function.  The points joined by a line are the data, the 
dashed line and grey band represents the mean difference and the 1 sigma 
fluctuation level seen in 2000 Monte Carlo simulations of random CMB 
skies.  Note that the FIRAS map has large-scale structure not seen in the DMR 
map (top panel) which is plausibly due to residual instrumental drifts 
corrupting the weak anisotropy.  This effect is largely uncorrelated with the 
DMR (middle and bottom panels).}
\label{2pt_fig}
\end{figure}

\clearpage

\begin{figure}[t]
\psfig{file=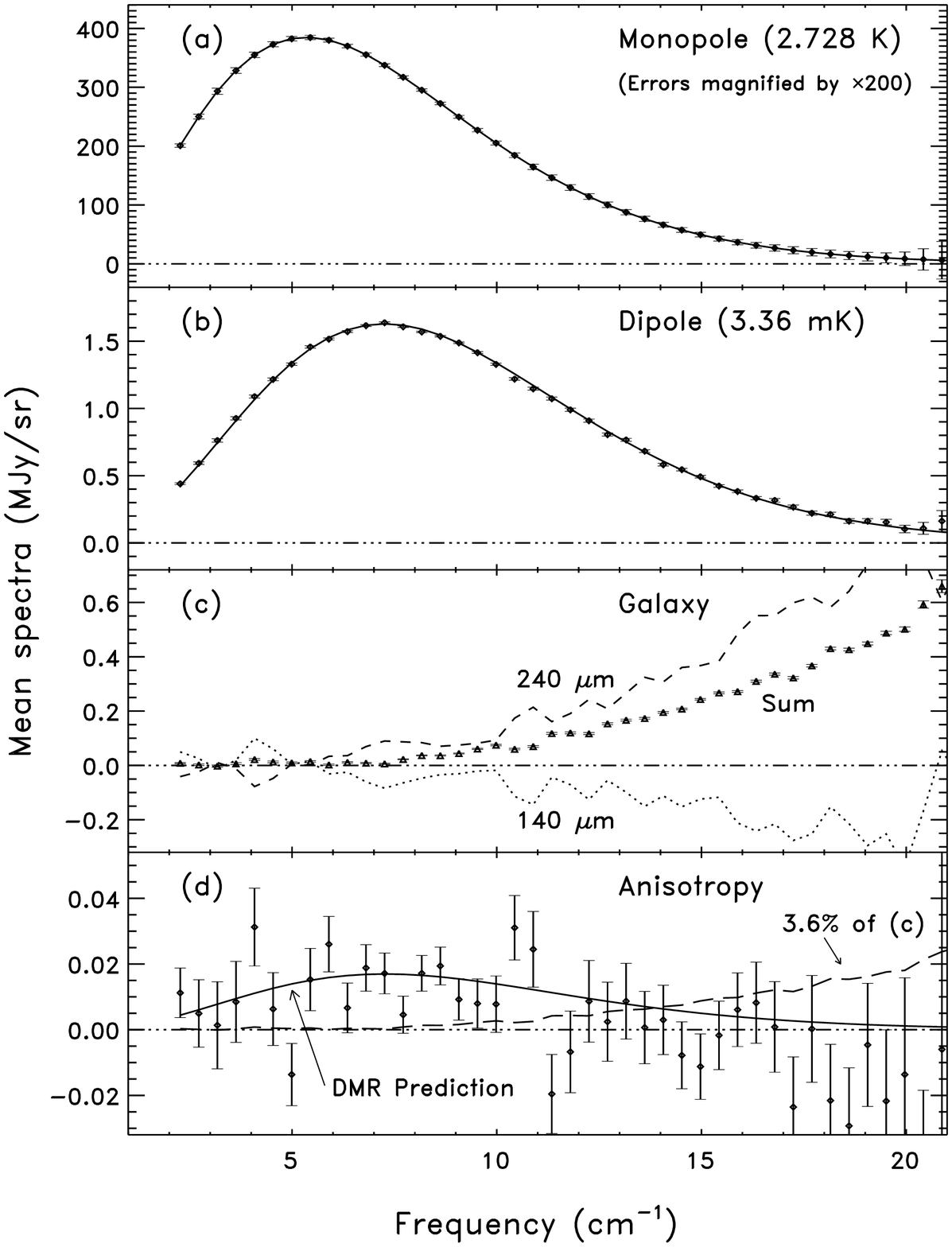,width=6.5in}
\end{figure}

\begin{figure}[t]
\psfig{file=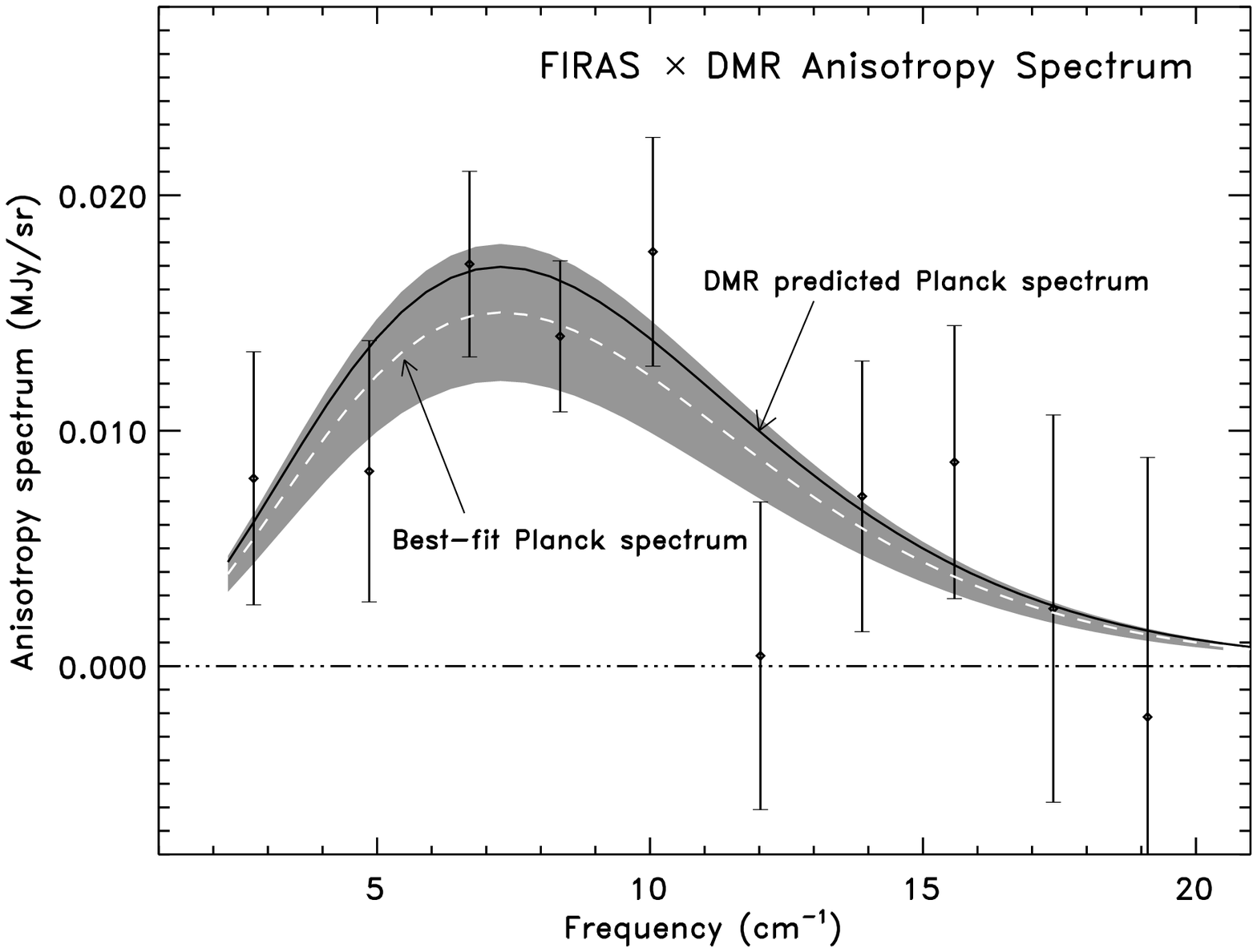,width=6.5in}
\end{figure}

\begin{figure}[t]
\psfig{file=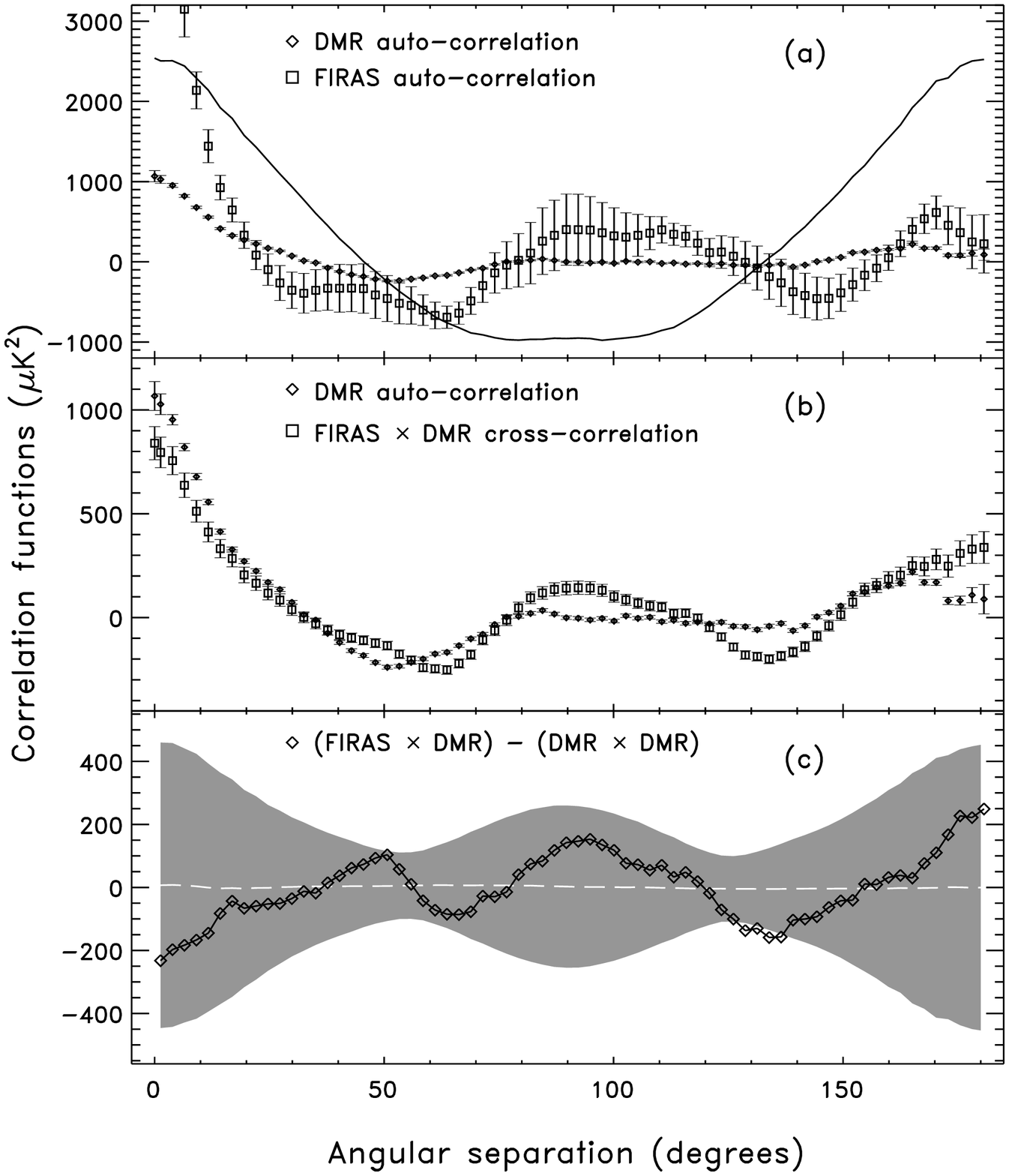,width=6.5in}
\end{figure}

\end{document}